\documentclass[prl,twocolumn,preprintnumbers]{revtex4}

\usepackage{amsmath}
\usepackage{graphicx}
\usepackage[small]{subfigure}
\usepackage[hyperfootnotes=false]{hyperref}

\newcommand{\eq}[1]{Eq.~\eqref{eq:#1}}
\newcommand{\eqs}[2]{Eqs.~\eqref{eq:#1} and \eqref{eq:#2}}
\newcommand{\fig}[1]{Fig.~\ref{fig:#1}}

\newcommand{\abs}[1]{\lvert#1\rvert}
\newcommand{\ord}[1]{\mathcal{O}(#1)}

\newcommand{\df}{\mathrm{d}}

\newcommand{\tr}{\mathrm{tr}}
\newcommand{\cI}{{\mathcal I}}
\newcommand{\hH}{\widehat{H}}
\newcommand{\hS}{\widehat{S}}

\newcommand{\nn}{\nonumber}

\newcommand{\Ecm}{E_\mathrm{cm}}

\allowdisplaybreaks[4]

\begin{document}


\preprint{\vbox{\hbox{MIT--CTP 4139}\hbox{April 14, 2010}}}

\title{\boldmath$N$-Jettiness: An Inclusive Event Shape to Veto Jets}

\author{Iain W.~Stewart}
\affiliation{Center for Theoretical Physics, Massachusetts Institute of
  Technology, Cambridge, MA 02139\vspace{-0.5ex}}

\author{Frank J.~Tackmann}
\affiliation{Center for Theoretical Physics, Massachusetts Institute of
  Technology, Cambridge, MA 02139\vspace{-0.5ex}}

\author{Wouter J.~Waalewijn\vspace{0.5ex}}
\affiliation{Center for Theoretical Physics, Massachusetts Institute of
Technology, Cambridge, MA 02139\vspace{-0.5ex}}

\begin{abstract}
  Jet vetoes are essential in many Higgs and new-physics analyses at the LHC and
  Tevatron.  The signals are typically characterized by a specific number of
  hard jets, leptons, or photons, while the backgrounds often have additional
  jets.  In such cases vetoing undesired additional jets is an effective way to
  discriminate signals and background.  Given an inclusive event sample with $N$
  or more jets, the veto to have only $N$ energetic jets defines an
  ``exclusive'' $N$-jet cross section.  This strongly restricts the phase space
  of the underlying inclusive $N$-jet cross section and causes large double
  logarithms in perturbation theory that must be summed to obtain theory
  predictions. Jet vetoes are typically implemented using jet algorithms.  This
  yields complicated phase-space restrictions and one often relies on
  parton-shower Monte Carlos, which are limited to leading-logarithmic accuracy.
  We introduce a global event shape ``$N$-jettiness'', $\tau_N$, which is
  defined for events with $N$ signal jets and vanishes in the limit of exactly
  $N$ infinitely narrow jets.  Requiring $\tau_N \ll 1$ constrains radiation
  between the $N$ signal jets and vetoes additional undesired jets.  This
  provides an inclusive method to veto jets and to define an exclusive $N$-jet
  cross section that can be well-controlled theoretically. $N$-jettiness yields
  a factorization formula with inclusive jet and beam functions.

\end{abstract}

\maketitle

\paragraph*{Introduction.}

At the LHC or Tevatron, hard interactions involving Higgs or new-physics particles
are identified by looking for signals with
a characteristic number of energetic jets, leptons, or photons~\cite{exp}.
The backgrounds come from Standard Model processes producing
the same signature of hard objects possibly with
additional jets. An example are top quarks decaying into $W$ plus
$b$-jet, which is a major background for $H\to WW$~\cite{Aaltonen:2010yv}.
When reconstructing masses and decay chains of new-physics particles
additional jets can cause large combinatorial backgrounds.
Standard Model processes can also fake a signal when a jet is
misidentified as lepton or photon, a typical example being $H\to \gamma\gamma$.

Thus, a veto on additional undesired jets is an effective and sometimes necessary method to
clean up the events and discriminate signal and the various backgrounds.
More generally, one would like to measure an ``exclusive''
$N$-jet cross section, $pp\to X L(Nj)$, to produce $N$ signal jets $j$
where the remaining $X$ contains no hard (central) jets.
Here, $N \geq 0$ and $L$ denotes the hard leptons or photons required as part of the signal.

We introduce an inclusive event shape ``$N$-jettiness'', denoted $\tau_N$ and
defined below in \eq{tauN}. For an event with at least $N$ energetic jets,
$\tau_N$ provides an inclusive measure of how $N$-jet-like the event looks. In
the limit $\tau_N \to 0$ the event contains exactly $N$ infinitely narrow jets.
For $\tau_N \sim 1$ the event has hard radiation between the $N$ signal jets.
Requiring $\tau_N \ll 1$ constrains the radiation outside the signal and beam
jets, providing an inclusive way to veto additional central jets. It yields an
inclusive definition of an exclusive $N$-jet cross section with a smooth
transition between the case of no jet veto, $\tau_N \sim 1$, and the extremely
exclusive case $\tau_N\to 0$.

Vetoing additional jets imposes a phase-space restriction on the underlying
inclusive $N$-jet cross section to produce $N$ or more jets with the same $L$.
Irrespective of its precise definition, the jet veto introduces a jet
resolution scale $\mu_J$ that characterizes this restriction, i.e.\ the
distinction between $N$ and $N\!+\!1$ jets. Hence, the exclusive $N$-jet cross
section contains phase-space logarithms $\alpha_s^n \ln^m(\mu_J^2/\mu_H^2)$,
where $m\leq 2n$ and $\mu_H$ is the scale of the hard interaction. For $\tau_N$,
$\mu_J^2/\mu_H^2 \simeq \tau_N \ll 1$. Generically there is always a hierarchy
$\mu_J \ll \mu_H$, which becomes larger the stronger the restrictions are. These
large logarithms must be summed to obtain reliable predictions.

Jet vetoes are typically implemented by using a jet algorithm to find all jets
in the event and veto events with too many energetic jets.  Jet algorithms are
good tools to identify the signal jets. However, they are not necessarily
well-suited to veto unwanted jets, because the corresponding phase-space
restrictions are complicated and depend in detail on the algorithm.  This makes
it difficult to incorporate the jet veto into explicit theoretical calculations
and inhibits a systematic summation of the resulting large logarithms.  In this
case, usually the only way to predict the corresponding exclusive $N$-jet cross
section is to rely on parton shower Monte Carlos to sum the leading logarithms
(LL). For particular jet algorithms, the resolution $y_{23}$ defines the
transition from $2$ to $3$ jets.  Next-to-leading logarithms for this and other
hadron-collider event shapes were summed in Ref.~\cite{Banfi:2004nk}.

Vetoing jets by cutting on an inclusive variable like $\tau_N$ has
several advantages. First, we can go beyond LL order, because the logarithms
from the phase-space restriction, $\alpha_s^n \ln^m\!\tau_N$, are simple enough
to allow their systematic summation to higher orders. Moreover, the theory
predictions with factorization can be directly compared with experiment without
having to utilize Monte Carlos for parton showering or hadronization.
Experimentally, $\tau_N$ reduces the dependence on jet algorithms and might help
improve the background rejection.

\begin{figure*}[ht!]
\hspace*{\fill}\subfigure[\hspace{1ex}$e^+e^- \to 2$ jets.]{%
\includegraphics[width=0.32\textwidth]{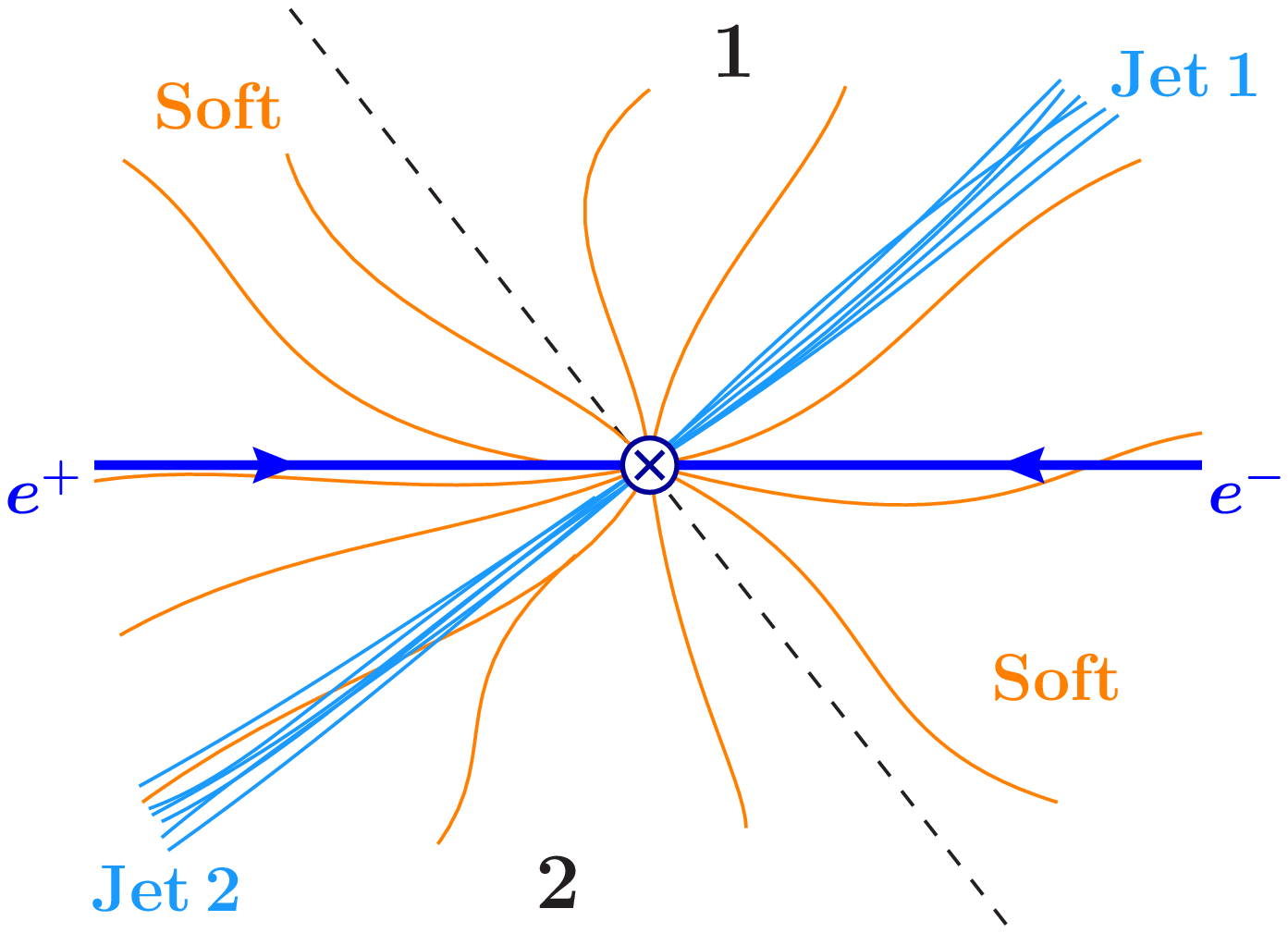}%
\label{fig:ee2jets}%
}\hfill%
\subfigure[\hspace{1ex}Isolated Drell-Yan.]{%
\includegraphics[width=0.32\textwidth]{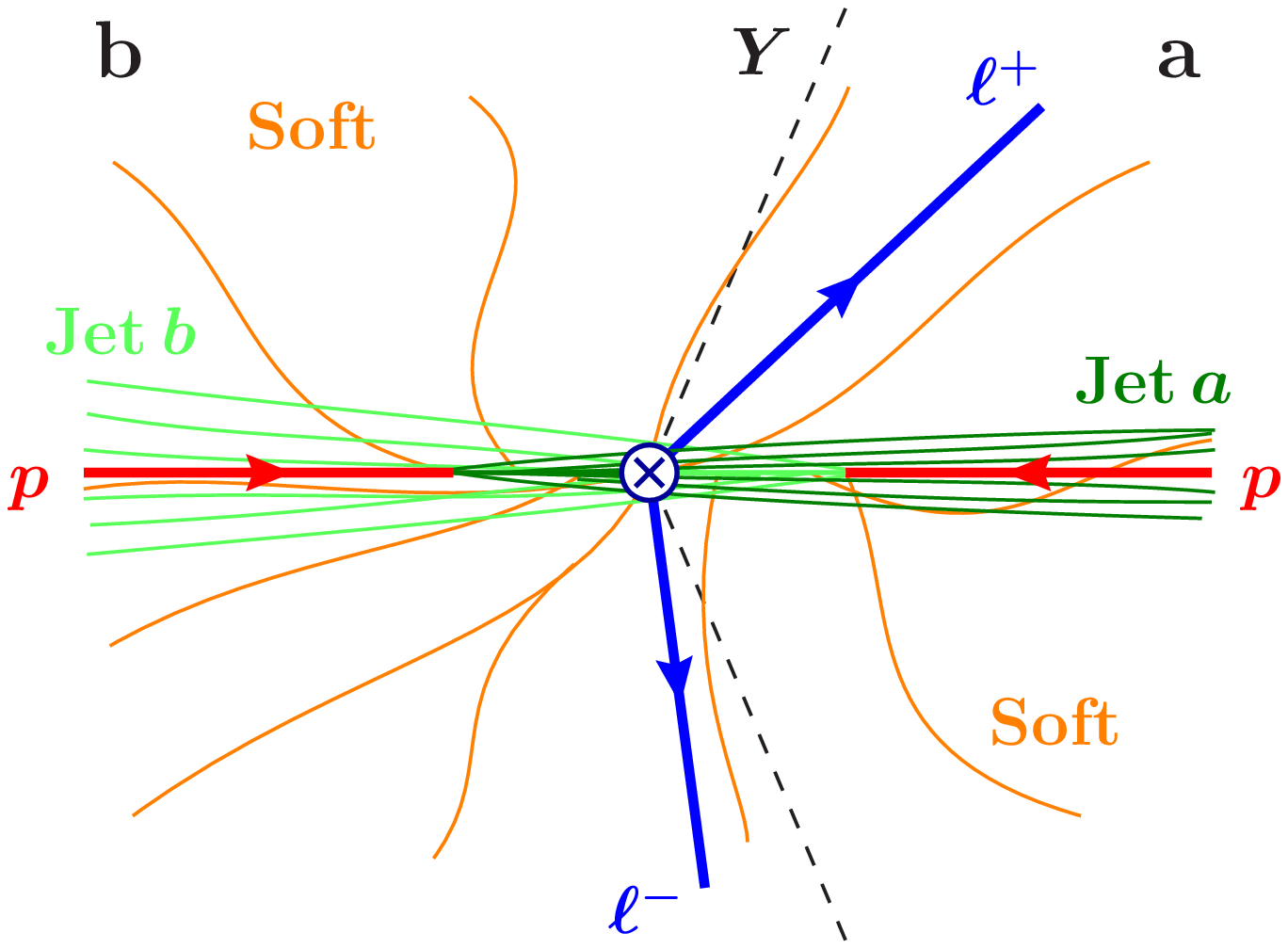}%
\label{fig:drellyan}%
}\hfill%
\subfigure[\hspace{1ex}$pp\to $ leptons plus jets.]{%
\includegraphics[width=0.32\textwidth]{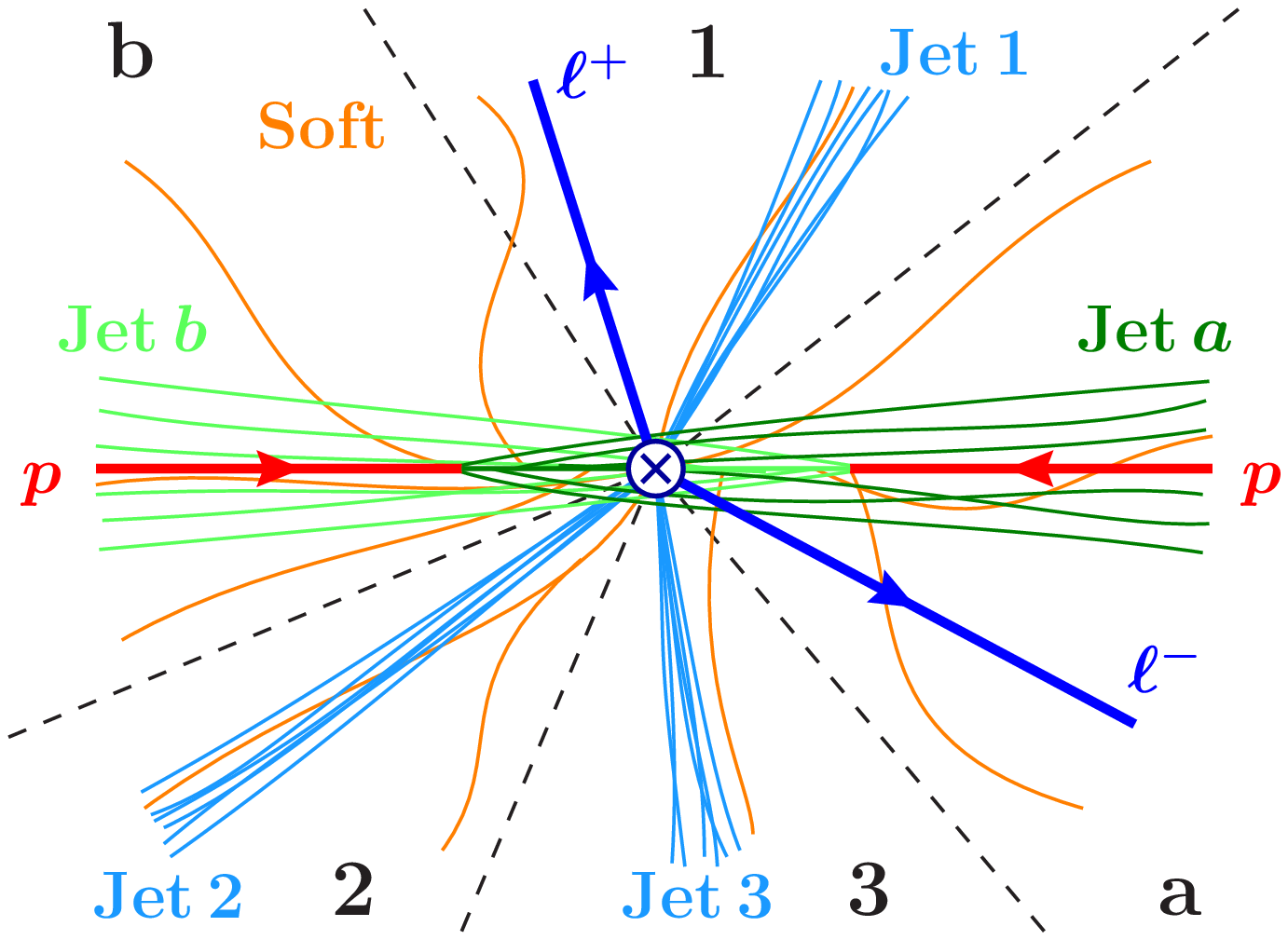}%
\label{fig:3jet}%
}\hfill%
\caption{\label{fig:pp} Different situations for the application of $N$-jettiness.}
\vspace{-1ex}
\end{figure*}

\paragraph*{Definition.}

$N$-jettiness is defined as
\begin{equation}\label{eq:tauN}
\tau_N = \frac{2}{Q^2}\sum_k
\min \bigl\{ q_a\cdot p_k,\, q_b\cdot p_k,\, q_1\cdot p_k,\, \ldots,\, q_N\cdot p_k \bigr\}
\,.\end{equation}
As we discuss below, this definition of $\tau_N$ yields a factorization formula
with inclusive jet and beam functions and allows the summation of logarithms to
next-to-next-to-leading logarithmic (NNLL) order.  The sum over $k$ in \eq{tauN}
runs over the momenta $p_k$ of all measured (pseudo-)particles in the final
state excluding the signal leptons or photons in $L$. (Any other leptons or
photons, e.g.\ from hadronic decays, are included in the sum.) For simplicity we
take all $p_k$ to be massless.  The $q_a$, $q_b$, and $q_1$, ..., $q_N$ are a
fixed set of massless reference momenta for the two beams and the $N$ signal
jets,
\begin{align}\label{eq:mom}
q_{a,b}^\mu &= \frac{1}{2}x_{a,b} \Ecm\, n_{a,b}^\mu
\,,& n_a^\mu &= (1, \hat{z})\,,\quad n_b^\mu = (1, -\hat{z})
\,,\nn\\
q_J^\mu &= E_J (1, \hat{n}_J)
\,,& J &= \{1,\ldots,N\}
\,.\end{align}
The $E_J$ and $\hat{n}_J$ correspond to the energies and directions of the $N$
signal jets (for both massive and massless jets). Their choice is discussed
below.  The beam reference momenta $q_a$ and $q_b$ are the large momentum
components of the colliding partons along the beam axis (taken to be the $z$
axis). They are defined by
\begin{equation}\label{eq:xab}
x_a \Ecm = n_b \cdot(q_1 + \dotsb + q_N + q)
\,,\end{equation}
and analogously for $x_b$ with $a\leftrightarrow b$. Here, $q$ is the total momentum of the non-hadronic signal $L$.
In \eq{tauN}, $Q^2 = x_a x_b\Ecm^2$ is the hard interaction scale, and
the distance of a particle with momentum $p_k$ from the jets or beams is measured by $q_m\cdot p_k$.
If $L$ contains missing energy, so $q$ and $x_{a,b}$ are not known,
one can use a modified distance measure as we discuss below \eq{geo}.

The minimum for each $k$ in \eq{tauN} associates the particle with the closest beam or jet,
appropriately dividing the hadronic initial-state radiation (ISR) and
final-state radiation (FSR). Soft particles and energetic particles near any jet
or beam only give small contributions to the sum. For $2\to N$ scattering of
massless partons, $\tau_N = 0$. Energetic particles far away from all jets and
beams give large contributions. Hence, for $\tau_N \ll 1$ the final state has
$N$ jets, two forward beam jets, and only soft radiation between them. In this
limit $x_{a,b}$ are the momentum fractions of the annihilated partons, and $Y =
\ln(x_a/x_b)/2$ is the boost of the partonic center-of-mass frame.

\paragraph*{$N = 2$ for $e^+e^-\to $ jets.}

In $e^+e^-$ collisions there is no hadronic ISR, so we drop the $q_{a,b}\cdot
p_k$ entries in \eq{tauN}. Now $Q^2$ is the total invariant mass of the leptons
and $Y = 0$. In the two-jet limit, the jet directions are close to the thrust
axis $\hat{t}$, defined by the thrust $T={\rm max}_{\hat t} \sum_i |\hat
t\cdot\vec p_i|/Q$. Hence we can choose
\begin{equation}
q_1^\mu = \frac{1}{2}\,Q\,(1,\hat{t}\,)
\,,\qquad
q_2^\mu = \frac{1}{2}\,Q\,(1,-\hat{t}\,)
\end{equation}
as reference momenta, and \eq{tauN} becomes
\begin{equation}\label{eq:tau2ee}
\tau^{ee}_2
= \frac{1}{Q} \sum_k E_k \min \bigl\{ 1-\cos\theta_k,\, 1 + \cos\theta_k \bigr\}
\,,\end{equation}
where $\theta_k$ is the angle between $\vec{p}_k$ and $\hat{t}$.  The minimum
divides all particles into the two hemispheres perpendicular to $\hat{t}$ as
shown in \fig{ee2jets}. For $\tau_2^{ee} \ll 1$, the total invariant mass in
each hemisphere is much smaller than $Q$, so the final state contains two narrow
jets. In this limit, $\tau_2^{ee} = 1 - T$, and a factorization theorem exists
for $\df\sigma/\df\tau_2^{ee}$, which can be used to sum logarithms of
$\tau_2^{ee}$~\cite{ee}. For a given jet algorithm with resolution parameter
$y$, the value $y_{23}$ marks the transition between $2$ and $3$ jets.
Thus requiring $y_{23} \ll 1$ also vetoes events with $> 2$ jets.

\paragraph*{$N = 0$ for Drell-Yan.}

Next, consider the isolated Drell-Yan process, $pp \to X\ell^+ \ell^-$ with no hard central jets, shown in \fig{drellyan}. We now have ISR from the incoming partons, but no FSR from jets. From \eq{xab} we have
\begin{equation} \label{eq:tau0_xab}
x_a \Ecm = e^{+Y}\! \sqrt{q^2 + \vec{q}_T^{\,2}}
\,,\quad
x_b \Ecm = e^{-Y}\! \sqrt{q^2 + \vec{q}_T^{\,2}}
\,,\end{equation}
where $q^2$ and $\vec{q}_T$ are the dilepton invariant mass and transverse momentum, and $Y$ equals the dilepton rapidity. Now, $Q^2 = q^2 + \vec{q}^{\,2}_T$ and \eq{tauN} becomes
\begin{equation} \label{eq:tau0}
\tau_0 = \frac{1}{Q} \sum_k \abs{\vec{p}_{kT}} \min\bigl\{e^{Y-\eta_k},\, e^{-Y+\eta_k} \bigr\}
\,.\end{equation}
where $\abs{\vec{p}_{kT}}$ and $\eta_k$ are the transverse momentum and rapidity
of $p_k$.  The $q_a$ and $q_b$ dependence in \eq{tauN} explicitly accounts
for the boost of the partonic center-of-mass frame. For $Y = 0$, the minimum in
\eq{tau0} divides all particles into two hemispheres perpendicular to the beam
axis (analogous to $\hat t$ above). For $Y \neq 0$, the hemispheres
are boosted with their boundary now at $Y$, and the beam jet in the direction of
the boost is narrower than the other, as depicted in \fig{drellyan}.
Contributions to $\tau_0$ from large rapidities are exponentially suppressed by
$\abs{\vec{p}_{kT}} e^{-\abs{\eta_k}} \approx 2E_k e^{-2\abs{\eta_k}}$, so
particles beyond the detector's rapidity reach give negligible contributions.

Beam thrust~\cite{Stewart:2009yx} is given by $\tau_B = \sqrt{1 + \vec{q}_T^{\,2}/q^2}\, \tau_0$. It is obtained by choosing $x_{a,b}\Ecm = \sqrt{q^2} e^{\pm Y}$ in case $q^2$ and $Y$ are measured rather than the longitudinal components $n_{a,b}\cdot q$ in \eq{tau0_xab}. For $\tau_0 \ll 1$ the hadronic final state can only contain soft radiation plus energetic radiation in the forward directions, so $\abs{\vec{q}_T} \ll Q$ and $\tau_B = \tau_0$. A factorization theorem for $\df\sigma/\df\tau_B$ at small $\tau_B$ was derived and used to sum logarithms of $\tau_B$ to NNLL in Refs.~\cite{Stewart:2009yx}.

\paragraph*{General case.}

For $pp\to XL (Nj)$ we have both ISR and FSR. We select candidate signal events
by measuring $L$ and running a jet algorithm to find the $N$ signal jets and
their momenta $p_J$. The conditions on the jets and $L$ that define the signal
are encoded in the cross section by a measurement function $F_N(\{p_J\}, L)$.
Generically, $F_N$ will enforce that there are at least $N$ energetic jets that
are sufficiently separated from each other and the beams. We now use the
measured jet energies and directions to define the massless reference
momenta $q_J$ in \eq{mom},
\begin{equation}
E_J = p_J^0
\,,\qquad
\hat{n}_J = \vec{p}_J/\abs{\vec{p}_J}
\,,\end{equation}
while $q_a$ and $q_b$ are given by \eqs{mom}{xab}.

Taking the minimum in \eq{tauN} combines the previous cases in \eqs{tau2ee}{tau0}. It divides all particles into jet and beam regions that are unique for a given set of reference momenta and whose union covers all of phase space, as illustrated in \fig{3jet}. The boundary between any two neighboring regions is part of a cone and is such that the sum of the total invariant masses in the regions is minimized
(or in case of a beam region the virtuality of the incoming colliding parton).

For events with small $\tau_N$ all jet algorithms should agree how energetic radiation is split up between the jets and beams, and only differ in their treatment of softer particles. Thus, they all give
the same $\hat{n}_J$ and $E_J$ up to power corrections, while the split up of the soft radiation is determined by $\tau_N$ itself. Hence, the dependence of $\tau_N$ on the jet algorithm is formally power suppressed, $\tau_N^\mathrm{alg. 1} = \tau_N^\mathrm{alg. 2} + \ord{\tau_N^2}$,
as seen in \eq{sigma} below.

To measure $\tau_N$, we still rely on having a suitable jet algorithm to find the $N$ signal jets but not more so than if we were not measuring $\tau_N$. Imagine the jet size in the algorithm is chosen too small such that the algorithm divides what should be a single signal jet into several narrow jets~\footnote{This can be tested by comparing the total energy in each region defined by $\tau_N$ with the energy from the jet algorithm. If these are very different, but at the same time $\tau_N$ is small, then there are additional energetic particles near the signal jets that the algorithm should have included.}.
In this case, the jet algorithm yields a poorly reconstructed signal irrespective of measuring $\tau_N$.

Since the jet veto is now provided by $\tau_N$, this situation can be avoided because we do not have to rely on the jet algorithm to identify additional jets and so can use an algorithm that can be forced to always yield at most $N$ jets. This is in fact the most natural thing to do when one is looking for $N$ jets. Therefore, using $\tau_N$ as jet veto could also help improve the signal reconstruction.

\paragraph*{Generalizations.}

We can generalize $\tau_N$ to
\begin{equation}
\tau_N^d = \sum_k \min \bigl\{ d_a(p_k), d_b(p_k), d_1(p_k),\ldots, d_N(p_k) \bigr\}
\,,\end{equation}
where $d_m(p_k)$ can be any infrared-safe distance measure. In \eq{tauN},
$d_m(p_k) = 2q_m\cdot p_k/Q^2$ with
\begin{align} \label{eq:qipk}
2q_a \cdot p_k &= \abs{\vec p_{kT}}\, Q\, e^{Y-\eta_k}
\,,\nn\\
2q_J \cdot p_k &= \abs{\vec p_{kT}}\, \abs{\vec q_{JT}}\, (2\cosh \Delta \eta_{Jk} - 2\cos \Delta \phi_{Jk})
\,.\end{align}
Here, $\Delta\eta_{Jk}$ and $\Delta\phi_{Jk}$ are the rapidity and azimuthal distances between $q_J$ and $p_k$. If these are small, the factor in brackets reduces to the familiar $R^2 = (\Delta \eta)^2 + (\Delta \phi)^2$.

Different measures that are boost-invariant along the beam axis can be obtained
by modifying the dependence on rapidity, $\abs{\vec q_{JT}}$, and $Q$ in
\eq{qipk}.  A geometric measure, which is independent of $\abs{\vec{q}_{JT}}$,
is
\begin{align} \label{eq:geo}
d_a(p_k) &= \frac{\abs{\vec p_{kT}}}{Q}\, e^{Y -\eta_k}
\,,\qquad
d_b(p_k) = \frac{\abs{\vec p_{kT}}}{Q}\, e^{-Y +\eta_k}
\,,\nn\\
d_J(p_k) &= \frac{\abs{\vec p_{kT}}}{Q}\, (2\cosh \Delta \eta_{Jk} - 2\cos \Delta \phi_{Jk})
\,.\end{align}
It evenly divides the area rather than invariant mass between neighboring regions, such that more energetic jets also get more invariant mass.

If $L$ contains missing energy, then $x_{a,b}$ in \eq{xab} and thus $Q$ and $Y$
are not known. For $Q$, one can use any hard scale, like the
$\abs{\vec{q}_{JT}}$ of the hardest jet or leave it out, since it only serves as
an overall normalization. In the beam measures $d_{a,b}(p_k)$ we can simply set
$Y = 0$, which defines them in the hadronic center-of-mass frame.

$N$-jettiness does not split events into $N$, $N+1$, $N+2$, etc. jets like a traditional jet algorithm.
But we can consider using $\tau_N$ to define an ``exclusive $N$-jet
algorithm'' as follows: First, we use a geometric measure and find the
directions $\hat{n}_J$ and boost $Y$ that minimize $\tau_N$, analogous
to finding $\hat t$ for $e^+ e^-\to {\rm jets}$.  This might actually allow
one to get an estimate of $Y$ even in the case of missing energy by exploiting
the asymmetry in the beam jets.  Second, we determine the jet energies by
summing over the particles in each jet region. (To reduce the sensitivity to the
underlying event and pile-up, one can weigh the sum over energies by the
distance from $\hat{n}_J$.)

\paragraph*{Factorization formula.}

We now use $\tau_N$ again as defined in \eq{tauN}.  For $\tau_N \ll 1$, QCD ISR
and FSR can be described in soft-collinear effective theory~\cite{SCET} at
leading power by $N+2$ independent sectors for collinear particles close to each
$q_m$ with $m = \{a,b,J\}$ and a separate sector for soft particles. By power
counting, $J$-collinear particles are closest to $q_J$, so for
the $J$-collinear sector
\begin{equation}
\sum_{k\in \mathrm{coll}_J} \min_m \bigl\{2q_m\cdot p_k \bigr\}
= \sum_{k\in \mathrm{coll}_J}  2q_J\cdot p_k = s_J
\,,\end{equation}
where (up to power corrections) $s_J$ is the total invariant mass in the $J$-collinear sector. Similarly, the sum over the beam collinear sectors yields the total (transverse) virtuality of the colliding partons, $t_a$ and $t_b$. Therefore,
\begin{equation} \label{eq:tauN_SCET}
\tau_N Q^2 = t_a + t_b + \sum_J s_J + \sum_{k\in\mathrm{soft}} \min_m \bigl\{2 q_m \cdot p_k \bigr\}
\,.\end{equation}
The sum in the last term is now restricted to the soft sector.  Combining
\eq{tauN_SCET} with the analyses in Refs.~\cite{Bauer:2008jx, Stewart:2009yx}
yields the factorization formula for $N$-jettiness \footnote{Here, $F_N$
  enforces distinct collinear sectors with $1-\hat n_l\cdot\hat n_m \gg \tau_N$
  and $E_m/Q \gg \tau_N$. We assume $F_N$ only depends on the large components
  $q_J$ of the jet momenta, $p_J = q_J[1 + \ord{\tau_N}]$, and that $L$ only
  couples to the QCD subprocess via a hard interaction. We also assume that Glauber gluons
do not spoil this factorization.}
\begin{align} \label{eq:sigma}
\frac{\df\sigma}{\df \tau_N}
&=
\int\!\df x_a \df x_b \int\!\df^4 q\,\df\Phi_L(q) \int\! \df \Phi_N(\{q_J\})
\nn\\ &\quad \times
F_N(\{q_m\}, L)\, (2\pi)^4 \delta^4\Bigl(q_a + q_b - \sum_J q_J - q\Bigr)
\nn\\ &\quad \times
\sum_{ij,\kappa} \tr\, \hH_{ij\to\kappa}(\{q_m\}, L, \mu) \prod_J \int\!\df s_J\, J_{\kappa_J}(s_J, \mu)
\nn\\ &\quad \times
\int\!\df t_a\, B_i(t_a, x_a, \mu)
\int\!\df t_b\, B_j(t_b, x_b, \mu)
\nn\\ &\quad \times
\hS_N^{ij\to\kappa} \Bigl(\tau_N - \frac{t_a + t_b + \sum_J s_J}{Q^2}, \{q_m\}, \mu\Bigr)
\,.\end{align}
Here, $\hH_{ij\to\kappa}(\{q_m\}, L)$ contains the underlying hard interaction $i(q_a)j(q_b) \to L(q) \kappa_1(q_1) \dotsb \kappa_N(q_N)$, where $i$, $j$, and $\kappa_J$ denote parton types, and the sum over $ij,\kappa$ is over all relevant partonic channels. It is a matrix in color space given by the IR-finite parts (in pure dim.\ reg.) of the squared partonic matrix elements in each channel. The $N$-body phase space for the massless momenta $q_J$ is denoted $\df\Phi_n(\{q_J\})$, and that for $L$ by $\df\Phi_L(q)$.

The inclusive jet and beam functions, $J_{\kappa_J}(s_J)$ and $B_{i,j}(t_{a,b},
x_{a,b})$, describe the final and initial state radiation emitted by the
outgoing and incoming partons from the hard interaction.  The latter also
determine the momentum fractions $x_{a,b}$ of the colliding partons and are
given by parton distribution functions $f_{i'}(\xi, \mu)$
as~\cite{Fleming:2006cd, Stewart:2009yx}
\begin{equation} \label{eq:B_fact}
B_i(t, x, \mu)
= \sum_{i'}\!\int_{x}^1 \frac{\df\xi}{\xi}\,
 \cI_{ii'}\Bigl(t,\frac{x}{\xi},\mu \Bigr) f_{i'}(\xi, \mu)
\,.\end{equation}
The $\cI_{ii'}$ are perturbative coefficients describing collinear ISR, and at
tree level $B_i(t, x, \mu) = \delta(t) f_i(x, \mu)$. The last term in \eq{tauN_SCET} is the contribution to $\tau_N$ from soft particles in the underlying event. It is described by the soft function $\hS_N^{ij\to\kappa}(\tau_N^\mathrm{soft}, \{q_m\})$, which depends on the jet's angles $\hat{n}_l\cdot \hat{n}_m$ and energy fractions $E_l/E_m$. Like $\hH$, it is a color matrix, and the trace in \eq{sigma} is over $\tr(\hH \hS)$.

In \eq{sigma}, all functions are evaluated at the same renormalization scale
$\mu$. Large logarithms of $\tau_N$ in $\df\sigma/\df\tau_N$ are summed by first
computing $\hH(\mu_H)$, $J(\mu_J)$, $B(\mu_B)$, $\hS(\mu_S)$ at the scales
$\mu_H \simeq Q$, $\mu_J\simeq\mu_B\simeq \sqrt{\tau_N}Q$, $\mu_S\simeq \tau_N
Q$, where the functions contain no large logarithms, and then evolving them to
the scale $\mu$.  This evolution is known analytically~\cite{evolution} and the
required anomalous dimensions are already known to NNLL~\cite{NNLL,
  Stewart:2009yx}, because we have inclusive jet and beam functions. NNLL also
requires the $\ord{\alpha_s}$ corrections for each function, which are known for
$J$ and $B$. The $\ord{\alpha_s}$ hard function is determined by the one-loop
QCD matrix elements. For $\tau_N \gg \Lambda_\mathrm{QCD}/Q$, $\hS(\mu_S)$ can
be computed perturbatively and will be given in a future publication.


We thank C.~Lee, K.~Tackmann, and J.~Thaler for comments and discussions.
This work was supported by the Office of Nuclear Physics of the U.S.\ Department
of Energy, under the grant DE-FG02-94ER40818.


\end{document}